%% file: lmcs2007.tex
\documentclass{LMCS}
\usepackage{hyperref}

\input{lambda}

\theoremstyle{plain}
\def\eg{{\em e.g.}}

\def\doi{3 (4:2) 2007}
\lmcsheading%
{\doi}
{1--8}
{}
{}
{Apr.~10, 2007}
{Nov.~\phantom{0}5, 2007}
{}
 
\begin{document}

\title[Shortest Developments]{A Note on Shortest Developments}

\author[M.~H.~S{\o}rensen]{Morten Heine S{\o}rensen}   %required
\address{Formalit, Byenden 32, 4660 Store Heddinge} %required
\email{mhs@formalit.dk}  %optional

%% etc.

%% required for running head on odd and even pages, use suitable
%% abbreviations in case of long titles and many authors:

%% mandatory lists of keywords and classifications:

%%%%%%%%%%%%%%%%%%%%%%%%%%%%%%%%%%%%%%%%%%%%%%%%%%%%%%%%%%%%%%%%%%%%%%%%%%%

%% the abstract has to PRECEED the command \maketitle:
%% be sure not to issue the \maketitle command twice!

\begin{abstract}
  \noindent De Vrijer has presented a proof of the {\em finite
developments\/} theorem which, in addition to showing that all
developments are finite, gives an effective reduction strategy
computing longest developments as well as a simple formula computing
their length.

We show that by applying a rather simple and intuitive principle of
duality to de Vrijer's approach one arrives at a proof that {\em
some\/} developments are finite which in addition yields an
effective reduction strategy computing shortest developments as well
as a simple formula computing their length. The duality fails for
general $\beta$-reduction.

Our results simplify previous work by Khasidashvili.

\end{abstract}

\maketitle

\section{Introduction}
Let $S = \{$ ``('', ``)'', ``.'', ``\underline{$\lambda$}'',
``{$\lambda$}'' $\}$ and $V$ be an infinite alphabet (ranged over by
$x,y,z$) disjoint from $S$. Then $\Lamlk$ is the set of words over
$S \cup V$ defined by:
$$\begin{array}{lcl}
 x \in V    & \mim & x \in \Lamlk \\
 P   \in \Lamlk & \mim & (\lam{x}{P}) \in \Lamlk \\
 P,Q \in \Lamlk & \mim &(\app{P}{Q}) \in \Lamlk \\
 P,Q \in \Lamlk & \mim & (\lbred{x}{P}{Q}) \in \Lamlk \\
\end{array}$$

We assume the reader is familiar with the fundamental conventions,
definitions, and properties pertaining to $\Lamlk$---see, \eg,
~\cite{barh84}---notably the conventions for omitting parentheses, the
notions of free and bound variables, the identification of terms
that differ only in the choice of names for bound variables, the
conventions for avoiding confusion between free and bound variables,
the definition of substitution $M\wth{x}{N}$, and the set $\FV(M)$ of
variables occurring free in $M$. Also, $M \equiv N$ means that $M$ and $N$
differ only in the choice of names for bound variables.

Let $\pil_\betal$ be the smallest relation on $\Lamlk$ with
$\lbred{x}{P}{Q} \pil_\betal P\wth{x}{Q}$ satisfying
$$\begin{array}{lcl}
 P \pil_\betal P' & \mim & \lam{x}{P} \pil_\betal \lam{x}{P'} \\
 P \pil_\betal P' & \mim & \app{P}{Q} \pil_\betal \app{P'}{Q} \\
 P \pil_\betal P' & \mim & \app{Q}{P} \pil_\betal \app{Q}{P'} \\
 P \pil_\betal P' & \mim & \bredl{x}{P}{Q} \pil_\betal \bredl{x}{P'}{Q} \\
 P \pil_\betal P' & \mim & \bredl{x}{Q}{P} \pil_\betal \bredl{x}{Q}{P'}
\end{array}$$

A development of $M_0$ is a finite or infinite sequence $M_0
\pil_\betal M_1 \pil_\betal \ldots$. If the sequence is finite, it
ends in the last term $M_n$ and has length $n$. If it is infinite, it
has length $\infty$.%
\footnote{We adopt the conventions $n \leq \infty$ and $\infty + n = \infty$ for all $n \in \nat \cup \{ \infty \}$.}
We write $M \in \NF_\betal$ and call $M$ a
$\betal$-normal form if $M \not\pil_\betal N$ for all $N\in\Lamlk$. A
development is complete if it is infinite or ends in a $\betal$-normal
form.  By $s_\betal(M)$ and $l_\betal(M)$ we denote the length of a
shortest complete and longest complete development of $M$,
respectively. The {\em finite developments\/} theorem, due to Curry and
Feys~\cite{curh58} and later proved by many others, states in its
simplest form that all developments are finite.

\begin{lem}\label{lem:devcalc}
\begin{thmcases}
\item $M, N \in \Lamlk \mim M\wth{x}{N} \in \Lamlk$;
\item $M \in \Lamlk \mco M \pil_\betal N \mim N \in \Lamlk$.
\end{thmcases}
\end{lem}
\proof (i): By induction on $M$. (ii): By induction on $M \pil_\betal
N$, using (i).\qed

\section{Shortest developments}
We first present our technique for computing shortest developments and
then explain the relation to de Vrijer's~\cite{vrir85} technique for
computing longest developments in \S 4.

\begin{defi}\label{def:mhto}
\begin{thmcases}
\item For all $x \in V$ define $m_x: \Lamlk \pil \nat$ by:%
\footnote{$\mn{m}{n}$ and $\mx{m}{n}$ denote the minimum and maximum of $m$ and $n$, respectively.}
$$\begin{array}{lcll}
m_x(x) & = & 1 \\
m_x(y) & = & 0 & \mbox{if $x \not\equiv y$}\\
m_x(\bredl{y}{P}{Q}) & = & m_x(P)+m_x(Q) \mn{m_y(P)}{1} \\
m_x(\app{P}{Q}) & = & m_x(P)+m_x(Q) & \mbox{if $P \not\equiv \laml{y}{R}$}\\
m_x(\lam{y}{P}) & = & m_x(P)
\end{array}$$

\item Define $h: \Lamlk \pil \nat$ by:
$$\begin{array}{lcll}
h(x) & = & 0 \\
h(\bredl{y}{P}{Q}) & = & h(P)+h(Q) \mn{m_y(P)}{1} +1 \\
h(\app{P}{Q}) & = & h(P)+h(Q) & \mbox{if $P \not\equiv \laml{y}{R}$}\\
h(\lam{y}{P}) & = & h(P)
\end{array}$$

\item Define $H: \Lamlk \pil \Lamlk$ by:
$$\begin{array}{lcll}
\stm{x} & = & x \\
\stm{\bredl{y}{P}{Q}} & = & \left\{ \begin{array}{l}
                                    \bredl{y}{P}{\stm{Q}}\\
                                    P\wth{y}{Q}
                 \end{array} \right. &
\begin{array}{l}
 \mbox{if $\mn{m_y(P)}{1}=1 \mco Q\not\in\NF_\betal$} \\
 \mbox{otherwise}
\end{array} \\
\stm{\app{P}{Q}} & = & \left\{ \begin{array}{l}
                    \app{\stm{P}}{Q} \\
                    \app{P}{\stm{Q}}
                \end{array}\right. &
\begin{array}{l}
 \mbox{if $P \not\equiv \laml{y}{R} \mco P \not\in\NF_\betal$}\\
 \mbox{if $P \not\equiv \laml{y}{R} \mco P \in \NF_\betal$}
\end{array} \\
\stm{\lam{y}{P}} & = & \lam{y}{\stm{P}}
\end{array}$$
\item Let $L_H(M)$ be the length of the complete development
$$M \pil_\betal H(M) \pil_\betal H(H(M)) \pil_\betal \ldots$$
\end{thmcases}
\end{defi}

As will be seen in Corollary~\ref{cor:mainto}, $M \pil_\betal H(M)
\pil_\betal H(H(M)) \pil_\betal \ldots$ is a shortest complete
development from $M$, and $h(M)$ is its length. Informally, the
auxiliary function $m_x(M)$ calculates the number of copies of $N$
we have to reduce in a shortest complete development of $M\{x:=N\}$.

\begin{rem}\label{rem:zeroto}
\begin{thmcases}
\item $x \not\in \FV(M) \mim m_x(M) = 0$.
\item $M \in \NF_{\betal} \mbi h(M) = 0$.
\item $\mn{m_y(P)}{1} \neq 1 \mim \mn{m_y(P)}{1} = m_y(P)$.
\end{thmcases}
\end{rem}

\begin{lem}\label{lem:multo}
Let $x \not\equiv y$. Then:
\begin{thmcases}
\item $m_y(M\wth{x}{N}) = m_y(M) + m_y(N)   m_x(M)$;
\item $h(M\wth{x}{N})   = h(M) + h(N)   m_x(M)$.
\end{thmcases}
\end{lem}
\proof (i) is by induction on $M$. Let $L^* \equiv L\wth{x}{N}$.
\begin{prfcases}
\item $M \equiv z$.
\begin{subcases}
\item $z \equiv x$. Then
$$\begin{array}{lcl}
m_y (x^*) & = & m_y(N) \\
          & = & m_y(x)+m_y(N)  m_x(x)
\end{array}$$

\item $z \not\equiv x$. Then
$$\begin{array}{lcl}
m_y(z^*) & = & m_y(z) \\
         & = &  m_y(z)+m_y(N)  m_x(z)
\end{array}$$
\end{subcases}

\item $M \equiv \bredl{z}{P}{Q}$. Since $z \not\in\FV(N)$, also $m_z(N)=0$.
By the induction hypothesis,
$$\begin{array}{l}
{m_y(\bredl{z}{P^*}{Q^*})} \\
 =  m_y(P^*)+m_y(Q^*)  \mn{m_z(P^*)}{1} \\
 =  m_y(P) + m_y(N)  m_x(P) + (m_y(Q) + m_y(N)  m_x(Q))\mn{m_z(P)}{1}\\
 =  m_y(P) + m_y(N)  m_x(P) + m_y(Q) \mn{m_z(P)}{1}+ m_y(N)m_x(Q)\mn{m_z(P)}{1} \\
 =  m_y(P)+m_y(Q)  \mn{m_z(P)}{1} + m_y(N)(m_x(P)+m_x(Q)  \mn{m_z(P)}{1})\\
 =  m_y(\bredl{z}{P}{Q}) + m_y(N)  m_x(\bredl{z}{P}{Q})
\end{array}$$

\item $M \equiv \app{P}{Q}$ where $P \not\equiv \laml{y}{R}$. Then, by the induction hypothesis,
$$\begin{array}{lcl}
m_y(\app{P^*}{Q^*}) & = & m_y(P^*)+m_y(Q^*) \\
 & = &  m_y(P) + m_y(N)  m_x(P) + m_y(Q) + m_y(N)  m_x(Q) \\
 & = &  m_y(\app{P}{Q}) + m_y(N)  m_x(\app{P}{Q})
\end{array}$$

\item $M \equiv \lam{y}{P}$. Similar to Case~3.

\end{prfcases}
This concludes the proof of (i); (ii) is also by induction on $M$.
\begin{prfcases}
\item $M \equiv z$.
\begin{subcases}
\item $z \equiv x$. Then
$$\begin{array}{lcl}
h(x^*) & = & h(N) \\
       & = & h(x)+h(N)   m_x(x)
\end{array}$$

\item $z \not\equiv x$. Then
$$\begin{array}{lcl}
h(z^*) & = & h(z) \\
       & = & h(z) + h(N)   m_x(z)
\end{array}$$
\end{subcases}

\item $M \equiv \bredl{z}{P}{Q}$. Since $z \not\in\FV(N)$, also $m_z(N)=0$.
Therefore, by the induction hypothesis and (i),
$$\begin{array}{ll}
\multicolumn{2}{l}{h(\bredl{z}{P^*}{Q^*})} \\
  = &  h(P^*)+h(Q^*)  \mn{m_z(P^*)}{1} + 1\\
  = &  h(P) +h(N)  m_x(P) + (h(Q)+h(N)  m_x(Q))\mn{m_z(P)}{1} +1 \\
  = &  h(P) + h(N)  m_x(P) + h(Q)   \mn{m_z(P)}{1} + h(N)m_x(Q)\mn{m_z(P)}{1} +1 \\
  = &  h(P)+h(Q)  \mn{m_z(P)}{1} + 1 + h(N)(m_x(P)+m_x(Q)  \mn{m_z(P)}{1})\\
  = &  h(\bredl{z}{P}{Q}) + h(N)  m_x(\bredl{z}{P}{Q})
\end{array}$$

\item $M \equiv \app{P}{Q}$ where $P \not\equiv \laml{y}{R}$. Then, by
the induction hypothesis,
$$\begin{array}{lcl}
h(\app{P^*}{Q^*}) & = & h(P^*)+h(Q^*) \\
 & = &  h(P) + h(N)  m_x(P) + h(Q) + h(N)  m_x(Q) \\
 & = &  h(\app{P}{Q}) + h(N)  m_x(\app{P}{Q})
\end{array}$$

\item $M \equiv \lam{y}{P}$. Similar to Case~3.\qed
\end{prfcases}

\begin{lem}\label{lem:decto}
Suppose that $M \pilbl N$. Then
\begin{thmcases}
\item $m_x(M) \leq m_x(N)$;
\item $h(M) \leq h(N) + 1$.
\end{thmcases}
\end{lem}
\proof (i) is by induction on $M \pilbl N$.
\begin{prfcases}
\item $M \equiv \bredl{y}{P}{Q} \pilbl P\wth{y}{Q} \equiv N$. By Lemma~\ref{lem:multo},
$$\begin{array}{lcl}
m_x(\bredl{y}{P}{Q}) & = & m_x(P)+m_x(Q)  \mn{m_y(P)}{1}\\
                     & \leq & m_x(P)+m_x(Q)  m_y(P)\\
                     & = & m_x(P\wth{y}{Q})
\end{array}$$

\item $M \equiv \bredl{y}{P}{Q} \pilbl \bredl{y}{P'}{Q'} \equiv N$, where
$P \pilbl P'$ and $Q \equiv Q'$, or vice versa. By the induction hypothesis,
$$\begin{array}{lcl}
m_x(\bredl{y}{P}{Q}) & =   & m_x(P) + m_x(Q) \mn{m_y(P)}{1} \\
                     &\leq &  m_x(P') + m_x(Q') \mn{m_y(P')}{1} \\
                     & =   & m_x(\bredl{y}{P'}{Q'})
\end{array}$$

\item $M \equiv \app{P}{Q} \pilbl \app{P'}{Q'} \equiv N$, where $P \not\equiv \laml{y}{R}$,
and where $P \pilbl P'$ and $Q \equiv Q'$, or vice versa. Similar to Case~2.

\item $M \equiv \lam{y}{P} \pilbl \lam{y}{P'} \equiv N$, where
$P \pilbl P'$. Similar to Case~2.
\end{prfcases}

This concludes (i); (ii) is also by induction on $M \pilbl N$.
\begin{prfcases}
\item $M \equiv \bredl{y}{P}{Q} \pilbl P\wth{y}{Q} \equiv N$. By Lemma~\ref{lem:multo}
$$\begin{array}{lcl}
h(\bredl{y}{P}{Q}) & = & h(P)+h(Q)  \mn{m_y(P)}{1} +1\\
           & \leq & h(P)+h(Q) m_y(P) +1\\
                   & = & h(P\wth{y}{Q}) + 1
\end{array}$$

\item $M \equiv \bredl{y}{P}{Q} \pilbl \bredl{y}{P'}{Q'} \equiv N$, where
$P \pilbl P'$ and $Q \equiv Q'$, or vice versa. By the induction hypothesis and (i),
$$\begin{array}{lcl}
h(\bredl{y}{P}{Q}) & = & h(P)+h(Q)\mn{m_y(P)}{1} + 1\\
                & \leq & h(P')+h(Q')\mn{m_y(P')}{1} + 2\\
        & = & h(\bredl{y}{P'}{Q'})+1
\end{array}$$

\item $M \equiv \app{P}{Q} \pilbl \app{P'}{Q'} \equiv N$, where $P \not\equiv \laml{y}{R}$,
and where
$P \pilbl P'$ and $Q \equiv Q'$, or vice versa. Similar to Case~2.

\item $M \equiv \lam{y}{P} \pilbl \lam{y}{P'} \equiv N$, where
$P \pilbl P'$. Similar to Case~2.\qed
\end{prfcases}

\begin{cor}\label{cor:fdvrito}
For all $M \in \Lamlk$: $h(M) \leq s_{\betal}(M)$.
\end{cor}
\proof By induction on $h(M)$.
\begin{prfcases}
\item $h(M)=0$. Then $M \in \NF_\betal$, and then $s_{\betal}(M)=0$.
\item $h(M) \neq 0$. Then $M \not\in \NF_\betal$.
Let $M \pilbl N$ be such that $s_{\betal}(M)=s_{\betal}(N)+1$. By
Lemma~\ref{lem:decto}(ii) and the induction hypothesis,
$$\begin{array}[b]{lcl}
h(M) & \leq & h(N) + 1 \\
     & \leq & s_{\betal}(N) + 1 \\
     & = & s_{\betal}(M)
\end{array}\eqno{\qEd}$$
\end{prfcases}

\begin{lem}\label{lem:perpdevto}
If $h(M) \neq 0$ then $M \pilbl \stm{M}$ and \mbox{$h(M)=h(\stm{M})+1$}.
\end{lem}
\proof By induction on $M$. Assume $h(M) \neq 0$.
\begin{prfcases}
\item $M \equiv x$. This case is impossible since $h(x)=0$.
\item $M \equiv \bredl{y}{P}{Q}$.
\begin{subcases}
\item $\mn{m_y(P)}{1}=1$ and $Q \not\in\NF_\betal$. By the induction hypothesis,
$$\begin{array}{lcl}
h(\bredl{y}{P}{Q})& = & h(P) + h(Q)\mn{m_y(P)}{1} + 1 \\
              & = & h(P) + h(Q) + 1\\
              & = & h(P) + h(\stm{Q}) + 2\\
              & = & h(P) + h(\stm{Q})\mn{m_y(P)}{1} + 2\\
          & = & h(\bredl{y}{P}{\stm{Q}}) + 1\\
          & = & h(\stm{\bredl{y}{P}{Q}}) + 1
\end{array}$$
\item $\mn{m_y(P)}{1} \neq 1$ or $Q \in \NF_\betal$. By Lemma~\ref{lem:multo}
$$\begin{array}{lcl}
h(\bredl{y}{P}{Q}) & = & h(P) + h(Q)\mn{m_y(P)}{1} + 1 \\
               & = & h(P) + h(Q)m_y(P) + 1\\
           & = & h(P\wth{y}{Q}) + 1
\end{array}$$
\end{subcases}

\item $M \equiv \lam{y}{P}$. Then, by the induction hypothesis,
$$\begin{array}{lcl}
h(\lam{y}{P}) & = & h(P) \\
              & = & h(\stm{P}) + 1\\
              & = & h(\lam{y}{\stm{P}}) + 1\\
              & = & h(\stm{\lam{y}{P}}) + 1
\end{array}$$
\item $M \equiv\app{P}{Q}$. Similar to Case~3.\qed
\end{prfcases}

\begin{cor}\label{cor:exactto}
For all $M \in \Lamlk$: $h(M)=L_H(M)$.
\end{cor}
\proof By induction on $h(M)$.
\begin{prfcases}
\item $h(M)=0$. Then $M \in \NF_\betal$, and then $L_H(M)=0$.
\item $h(M) \neq 0$. Then $M \not\in \NF_{\betal}$, and then by
Lemma~\ref{lem:perpdevto} and the induction hypothesis,
$$\begin{array}[b]{lcl}
h(M) & = & h(\stm{M}) + 1 \\
     & = & L_H(\stm{M}) + 1 \\
     & = & L_H(M)
\end{array}\eqno{\qEd}$$
\end{prfcases}

\begin{cor}\label{cor:mainto}
For all $M \in \Lamlk$: $h(M)=s_{\betal}(M)=L_H(M)$.
\end{cor}
\proof Let $M \in \Lamlk$. Obviously, $s_\betal(M) \leq L_H(M)$. By
Corollary~\ref{cor:fdvrito} and~\ref{cor:exactto},
$$s_{\betal}(M) \leq L_H(M) = h(M) \leq s_{\betal}(M)\eqno{\qEd}$$

\section{Relation to Khasidashvili's technique}
Khasidashvili~\cite{khaz88a} calls a redex $\Delta$ in $M$ {\em essential}, notation $E(\Delta,M)$,
if every complete development of $M$ must reduce $\Delta$ (or a residual of $\Delta$).
He shows that any strategy which reduces in each step an inner-most essential redex yields shortest complete
developments, and he gives a formula for the length of such developments:
the number of essential redexes in the initial term.  He also gives an
algorithm to decide whether a redex in a term is essential; this makes
the above strategy and formula effective, but the algorithm is---in our
opinion---somewhat involved. The algorithm can be simpler formulated in terms
of the map $m_y$ as follows:
$$\begin{array}{l@{\mbi}l}
E(\Delta,\bredl{y}{P}{Q}) & \Delta \equiv \bred{y}{P}{Q} \mbox{ or } E(\Delta,P) \mbox{ or } [E(\Delta,Q) \mco m_y(P)>0]\\
E(\Delta,\app{P}{Q})      & E(\Delta,P) \mbox{ or } E(\Delta,Q)\\
E(\Delta,\lam{y}{P})     & E(\Delta,P)
\end{array}$$
In this terminology, the map $h$ counts the number of essential
redexes in a term, and $H$ reduces \emph{some} essential redex whose
argument does not contain another essential redex.

\section{Relation to de Vrijer's technique}
De Vrijer~\cite{vrir85} studies the following maps $n_x$, $g$, and $G$, which arise from
$m_x$, $h$, and $H$ by replacing all minimum operators $\mn{\bullet}{\bullet}$ by
maximum operators $\mx{\bullet}{\bullet}$; intuitively this makes sense
since we now consider longest instead of shortest developments.
\begin{thmcases}
\item For all $x\in V$ define $n_x: \Lamlk \pil \nat$ by:
$$\begin{array}{lcll}
n_x(x) & = & 1 \\
n_x(y) & = & 0 & \mbox{if $x \not\equiv y$}\\
n_x(\bredl{y}{P}{Q}) & = & n_x(P)+n_x(Q) \mx{n_y(P)}{1} \\
n_x(\app{P}{Q}) & = & n_x(P)+n_x(Q) & \mbox{if $P \not\equiv \laml{y}{R}$}\\
n_x(\lam{y}{P}) & = & n_x(P)
\end{array}$$

\item Define $g: \Lamlk \pil \nat$ by:
$$\begin{array}{lcll}
g(x) & = & 0 \\
g(\bredl{y}{P}{Q}) & = & g(P)+g(Q) \mx{n_y(P)}{1} +1 \\
g(\app{P}{Q}) & = & g(P)+g(Q) & \mbox{if $P \not\equiv \laml{y}{R}$}\\
g(\lam{y}{P}) & = & g(P)
\end{array}$$
\item Define $G: \Lamlk \pil \Lamlk$ by:
$$\begin{array}{lcll}
\sti{x} & = & x \\
\sti{\bredl{y}{P}{Q}} & = & \left\{ \begin{array}{l}
                                    \bredl{y}{P}{\sti{Q}}\\
                                    P\wth{y}{Q}
                 \end{array} \right. &
\begin{array}{l}
 \mbox{if $\mx{n_y(P)}{1}=1 \mco Q \not\in\NF_\betal$} \\
 \mbox{otherwise}
\end{array} \\
\sti{\app{P}{Q}} & = & \left\{ \begin{array}{l}
                    \app{\sti{P}}{Q} \\
                    \app{P}{\sti{Q}}
                \end{array}\right. &
\begin{array}{l}
 \mbox{if $P \not\equiv \laml{y}{R} \mco P \not\in\NF_\betal$}\\
 \mbox{if $P \not\equiv \laml{y}{R} \mco P \in\NF_\betal$}
\end{array} \\
\sti{\lam{y}{P}} & = & \lam{y}{\sti{P}}
\end{array}$$
\item Let $L_G(M)$ be the length of the complete development
$$M \pil_\betal G(M) \pil_\betal G(G(M)) \pil_\betal \ldots$$
\end{thmcases}

De Vrijer proves that
$M \pil_\betal G(M) \pil_\betal G(G(M)) \pil_\betal \ldots$
is a longest complete development from $M$, and that $g(M)$ is the length of this development.
This is expressed by the equations: $L_G(M)=l_\betal(M)=g(M)$.
The finite developments theorem is an immediate corollary.

The proof of these equations can be carried out {\em exactly\/} as in~\ref{rem:zeroto}--\ref{cor:mainto} by replacing
$s_\betal$, $\mn{\bullet}{\bullet}$, $\leq$, $m_x$, $h$, and $L_H$ by
$l_\betal$, $\mx{\bullet}{\bullet}$, $\geq$, $n_x$, $g$, and $L_G$, respectively!
This works because the properties used in~\ref{rem:zeroto}--\ref{cor:mainto} involving
$s_\betal, m_x$, \etc\ are invariant under the transformation, as the reader is encouraged to check.%
\footnote{To obtain this result, a small change has been made to $G$
as compared to de Vrijer's formulation; in his formulation the
condition $\mx{n_y(P)}{1}=1$ is $n_y(P)=0$---see \S 5.}  For instance, the
property $\mn{m}{n} \leq m$ becomes $\mx{m}{n} \geq m$.

\section{Discussion}
Although the general notions of longest and shortest complete
$\beta$-reduction sequences are intuitively ``opposite,'' they are,
technically speaking, very different. For instance, there is an
effective reduction strategy that computes longest complete
$\beta$-reduction sequences (see~\cite{sorm96} among others), but no
effective reduction strategy that computes shortest complete
$\beta$-reduction sequences~\cite{barh84}.  In contrast, the above
shows that one can effectively compute both shortest and longest
complete developments, and the proofs reveal a duality between the
two concepts. It is natural to ask why the duality does not carry
over to the general case of $\beta$-reduction.

The difference between the minimal strategy $H$ and the maximal
strategy $G$ is revealed on terms of form $\bredl{y}{P}{Q}$ where
$Q\not\in\NF_{\betal}$.  The rationale behind the minimal strategy is
that if all reductions of $\bredl{y}{P}{Q}$ to $\betal$-normal
form must reduce inside at least one residual of $Q$, then it is best to perform reductions in $Q$
first, to avoid proliferation. This is
decidable for developments, but undecidable for $\beta$-reduction~\cite{barh87}.

The rationale behind the maximal strategy is that if all reductions of
$\bredl{y}{P}{Q}$ to $\betal$-normal form may reduce inside at most
one residual of $Q$, then it is best to perform reductions in $Q$
first, to avoid erasing. An equivalent technique, used by de
Vrijer~\cite{vrir85}, is to test whether reducing $\bredl{y}{P}{Q}$
one step would delete $Q$, and if so reduce $Q$ to normal form
first. This is decidable for developments as well as for
$\beta$-reduction.

From the point of view of efficiency, a minimal strategy is clearly
better than a maximal strategy. It is a remarkable fact that in
general $\beta$-reductions we can effectively do the worst possible
job, but not the best possible job.%
\footnote{But see~\cite{oosv2007} for a technique to establish
both longest and shortest reductions, though they may not be
effective.}

\end{document}

%% file: lambda.tex
\newcommand{\lam}[2]{\lambda #1.#2}

\newcommand{\app}[2]{#1 \; #2}

\newcommand{\bred}[3]{\app{(\lam{#1}{#2})}{#3}}

\newcommand{\laml}[2]{\underline{\lambda}#1.#2}
\newcommand{\bredl}[3]{\app{(\laml{#1}{#2})}{#3}}

\newcommand{\betal}{{\underline{\beta}}}

\newcommand{\pilbl}{\pil_{\betal}}

\newcommand{\Lamlk}{\underline{\Lambda}_K}

\newcommand{\NF}{\hbox{\small\rm NF}}

\newcommand{\FV}{\mbox{\small\rm FV}}

\newcommand{\mim}{\:\:{\bf \Rightarrow}\:\:}
\newcommand{\mbi}{\:\:{\bf \Leftrightarrow}\:\:}

\newcommand{\mco}{\:\:{\bf \&}\:\:}

\newcommand{\pil}{ \rightarrow }

\newcommand{\mx}[2]{\lceil #1,#2\rceil}
\newcommand{\mn}[2]{\lfloor #1,#2\rfloor}

\newcommand{\eg}{e.g.}

\newcommand{\etc}{etc.}

\newcommand{\wth}[2]{\{#1:=#2\}}

\newcommand{\godown}{\mbox{ }}

\newcommand{\nat}{\mathbb N}

\newcommand{\ignore}[1]{}

{\godown\begin{enumerate}

}
{\end{enumerate}}

{\godown
\begin{enumerate}
\setlength{\leftmargin}{0cm}

\vspace{-0.2cm}}%
{\end{enumerate}}

\newcounter{thmcas}
\newcounter{prfcas}
\newcounter{subcas}
\newcounter{subsubcas}
\newcommand{\spc}{
\setlength{\labelwidth}{0.8cm}
\setlength{\itemindent}{0cm}
\setlength{\labelsep}{0.2cm}
\setlength{\leftmargin}{0.8cm}
\setlength{\parsep}{0.0cm}}
\newenvironment{thmcasesa}{\begin{list}{\rm (\roman{thmcas})}{\usecounter{thmcas}\spc}}{\end{list}}
\newenvironment{prfcasesa}{\begin{list}{\arabic{prfcas}.}{\usecounter{prfcas}\spc}}{\end{list}}
\newenvironment{subcasesa}{\begin{list}{\arabic{prfcas}.\arabic{subcas}.}{\usecounter{subcas}\spc}}{\end{list}}
\newenvironment{subsubcasesa}{\begin{list}{\arabic{prfcas}.\arabic{subcas}.\arabic{subsubcas}}{\usecounter{subsubcas}\spc}}{\end{list}}
\newenvironment{thmcases}{\godown\begin{thmcasesa}}{\end{thmcasesa}}
\newenvironment{prfcases}{\godown\begin{prfcasesa}}{\end{prfcasesa}}
\newenvironment{subcases}{\godown\begin{subcasesa}}{\end{subcasesa}}

\newcommand{\lbred}[3]{(\underline{\lambda}#1.#2)#3}

\newcommand{\stm}[1]{H(#1)}
\newcommand{\sti}[1]{G(#1)}

%% file: lmcs2007.bbl
\begin{thebibliography}{1}

\bibitem{barh84}
H.P. Barendregt.
\newblock {\em The Lambda Calculus: {I}ts Syntax and Semantics}.
\newblock North-Holland, second, revised edition, 1984.

\bibitem{barh87}
H.P Barendregt, J.R. Kennaway, J.W. Klop, and M.R. Sleep.
\newblock Needed reduction and spine strategies for the lambda calculus.
\newblock {\em Information and Computation}, 75(3):191--231, 1987.

\bibitem{curh58}
H.B. Curry and R.~Feys.
\newblock {\em Combinatory Logic}.
\newblock North-Holland, 1958.

\bibitem{khaz88a}
Z.~Khasidashvili.
\newblock $\beta$-reductions and $\beta$-developments with the least number of
  steps.
\newblock In P.~Martin-L\"of and G.~Mints, editors, {\em International
  Conference on Computer Logic}, volume 417 of {\em Lecture Notes in Computer
  Science}, pages 105--111. Springer-Verlag, 1988.

\bibitem{oosv2007}
V.~van Oostrom.
\newblock Random descent.
\newblock In F.~Baader, editor, {\em Rewriting Techniques and Applications},
  volume 4533 of {\em Lecture Notes in Computer Science}, pages 314--328.
  Springer-Verlag, 2007.

\bibitem{sorm96}
M.H. S{\o}rensen.
\newblock Effective longest and infinite reduction paths in untyped
  $\lambda$-calculi.
\newblock In H.~Kirchner, editor, {\em Colloquium on Trees in Algebra and
  Programming}, volume 1059 of {\em Lecture Notes in Computer Science}, pages
  287--301. Springer-Verlag, 1996.

\bibitem{vrir85}
R.C.~de Vrijer.
\newblock A direct proof of the finite developments theorem.
\newblock {\em Journal of Symbolic Logic}, 50:339--343, 1985.

\end{thebibliography}
